\documentclass[12pt]{article}
\usepackage[dvips]{graphicx}
\def\be{\begin{equation}}					 
\def\ee{\end{equation}}
\def\ber{\begin{eqnarray}}
\def\eer{\end{eqnarray}}	

\begin{document}
\vspace*{1cm}
\begin{center}
{\Large \bf Kaluza-Klein Picture and Nucleon-Nucleon \\[1ex]
Dynamics at Low Energies }\\

\vspace{4mm}

{\large A.A. Arkhipov\\
{\it State Research Center ``Institute for High Energy Physics" \\
 142280 Protvino, Moscow Region, Russia}}\\
\end{center}

\vspace{4mm}
\begin{abstract}
{In this note we present additional arguments in favour of
Kaluza and Klein picture of the world. We show that geniusly
simple formula provided by Kaluza-Klein approach gives an
excellent description for the mass spectrum of two-nucleon system.
It has also been established that the experimental data obtained at
low energies where the nucleon-nucleon dynamics has been studied
reveal a special sort of (super)symmetry between fermionic (dibaryon)
and bosonic states predicted by Kaluza-Klein scenario.} 
\end{abstract}

\section{Introduction}

In our previous paper \cite{1} we have presented the arguments in
favour of that the Kaluza-Klein picture of the world has been been
observed for a long time in the experiments at very low energies
where the nucleon-nucleon dynamics has been studied. Actually, we
have shown that the structure of proton-proton total cross section at
very low energies has a clear signature of the existence of the extra
dimensions: It was found that geniusly simple formula for KK
excitations  provided by Kaluza-Klein approach accurately described
the experimentally observed irregularities in the spectrum of
mass of the diproton system. Surely, this was a very nice fact and,
certainly, it was not an accidental coincidence. 

The central point of  Kaluza-Klein approach is related to the
existence of a new fundamental scale characterizing a size of compact
internal extra space. The detailed study of the structure for the
proton-proton total cross section at very low energies allowed us to
calculate this scale. It turned out the fundamental scale has a clear
physical meaning: This scale corresponds to the scale of distances
where the strong Yukawa forces in strength come down to the
electromagnetic forces. We have shown that the Kaluza-Klein tower of
KK excitations built by the calculated fundamental scale was in a
good correspondence with the experimentally observed picture of
irregularities in the mass spectrum of diproton system. However, it's
clear that Kaluza-Klein scenario predict the same KK excitations in
the proton-antiproton system as well, and this is a very nontrivial
fact. That is why, it was intriguing for us to find an experimental
confirmation of this fact. We have performed an analysis of
experimental data on mass spectrum of the resonance states of
proton-antiproton system above elastic threshold and compared them
with Kaluza-Klein picture. The results of this analysis have
been presented here.  

\section{Kaluza-Klein picture and KK excitations in two-nucleon
system}

It is well known that the basic idea of the Kaluza-Klein scenario may
be applied to any model in Quantum Field Theory (see for the details
e.g. the excellent review articles \cite{2,3} and many references
therein). As example, let us consider the simplest case of
(4+d)-dimensional model of scalar field with the action  
\be
S = \int d^{4+d}z \sqrt{-{\cal G}} \left[
\frac{1}{2} \left( \partial_{M} \Phi \right)^2 - 
\frac{m^{2}}{2} \Phi^2 + \frac{G_{(4+d)}}{4!} \Phi^4
\right], 
\label{S}
\ee
where ${\cal G}=\det|{\cal G}_{MN}|$, ${\cal G}_{MN}$ is the metric
on ${\cal M}_{(4+d)} = M_4 \times K_d$, $M_4$ is pseudo-Euclidean
Minkowski space-time, $K_d$ is a compact internal $d$-dimensional
space with the characteristic size $R$. Let $\Delta_{K_{d}}$ be the
Laplace operator on the internal space $K_{d}$, and $Y_{n}(y)$ are
ortho-normalized eigenfunctions of the Laplace operator 
\be
\Delta_{K_{d}} Y_{n}(y) = -\frac{\lambda_{n}}{R^{2}} Y_{n}(y),  
\label{Yn}
\ee
and $n$ is a (multi)index labeling the eigenvalue
$\lambda_{n}$ of the eigenfunction $Y_{n}(y)$. $d$-dimensional torus
$T^{d}$ with equal radii $R$ is an especially simple example of the
compact internal space of extra dimensions $K_d$. The eigenfunctions
and eigenvalues in this special case look like 
\be
Y_n(y) = \frac{1}{\sqrt{V_d}} \exp \left(i \sum_{m=1}^{d}
n_{m}y^{m}/R
\right), \label{T}
\ee
\[
\lambda_n = |n|^2,\quad |n|^2= n_1^2 + n_2^2 + \ldots n_d^2, \quad
n=(n_1,n_2, \ldots, n_d),\quad -\infty \leq n_m \leq \infty,
\]
where $n_m$ are integer numbers, $V_d = (2\pi R)^d$ is the
volume of the torus.

To reduce the multidimensional theory to the effective
four-dimensional one we wright a harmonic expansion for
the multidimensional field $\Phi(z)$ 
\be
\Phi(z) = \Phi(x,y) = \sum_{n} \phi^{(n)}(x) Y_{n}(y). 
\label{H}
\ee
The coefficients $\phi^{(n)}(x)$ of the harmonic expansion
(\ref{H}) are called Kaluza-Klein (KK) excitations or KK modes, and
they usually include the zero-mode $\phi^{(0)}(x)$, corresponding to
$n=0$ and the eigenvalue $\lambda_{0} = 0$. Substitution of the KK
mode expansion into action (\ref{S}) and integration over the
internal space $K_{d}$ gives
\be
S = \int d^{4}x \sqrt{-g} \left\{  
\frac{1}{2} \left( \partial_{\mu} \phi^{(0)} \right)^{2} -
\frac{m^{2}}{2}
(\phi^{(0)})^{2} \right. + \frac{g}{4!} (\phi^{(0)})^{4} +
\ee
\[
+\left. \sum_{n \neq 0} \left[\frac{1}{2}
\left(\partial_{\mu} \phi^{(n)}
\right) 
\left(\partial^{\mu} \phi^{(n)} \right)^{*} -\frac 
{m_n^2}{2} \phi^{(n)}\phi^{(n)*} \right] 
+ \frac{g}{4!} (\phi^{(0)})^{2} \sum_{n\neq 0} \phi^{(n)}
\phi^{(n)*}\right\} + \ldots.  
\]
For the masses of the KK modes one obtains 
\be
m_{n}^{2} = m^{2} + \frac{\lambda_{n}}{R^{2}}, \label{m}
\ee
and the coupling constant $g$ of the four-dimensional theory is
related  to the coupling constant $G_{(4+d)}$ of the initial
multidimensional theory by the equation 
\be
  g = \frac{G_{(4+d)}}{V_d},  \label{g}
\ee
where $V_d$ is the volume of the compact internal space of extra
dimensions $K_d$. The fundamental coupling constant $G_{(4+d)}$ has
dimension $[\mbox{mass}]^{-d}$. So, the four-dimensional coupling
constant $g$ is dimensionless one as it should be.
Eqs.~(\ref{m},\ref{g}) represent the basic
relations of Kaluza-Klein scenario.  Similar relations take place for
other types of multidimensional quantum field theoretical models.
From four-dimensional point of view we can interpret each KK mode as
a particle with the mass $m_n$ given by Eq.~(\ref{m}). We see that in
according with Kaluza-Klein scenario any multidimensional
field contains an infinite set of KK modes, i.e. an infinite set of
four-dimensional particles with increasing masses, which is called
the Kaluza-Klein tower. Therefore, an experimental observation of
series KK excitations with a characteristic spectrum of the form
(\ref{m}) would be an evidence of the existence of extra dimensions.

As it was mentioned above in Introduction we have applied the main
issues of Kaluza-Klein approach to our analysis of the structure of
proton-proton total cross section at very low energies and calculated
the fundamental scale (size) $R$ of the compact internal extra space.
One obtained by this way
\be
\frac{1}{R} = 41.481\,\mbox{MeV}\quad \mbox{or}\quad
R=24.1\,GeV^{-1}=4.75\,10^{-13}\mbox{cm}.\label{scale}
\ee
After that we have built the Kaluza-Klein tower of KK
excitations by the formula 
\be
M_n=2\sqrt{m_p^2+\frac{n^2}{R^2}},\quad (n=1,2,3,\ldots)\label{KK}
\ee
and compared it with the observed irregularities in the spectrum of
mass of the diproton system. The result of the comparison has been
presented in Table 2 of ref. \cite{1}. Now we would like to include
in this table the experimental data on mass spectrum of the resonance
states of proton-antiproton system above elastic threshold taking
into account that Kaluza-Klein scenario predict $M_n^{pp}=M_n^{p\bar
p}$. The extended version of Table 2 from ref. \cite{1} is shown here
in Table 1. 

\begin{table}[ht]
\begin{center}
\caption{Kaluza-Klein tower of KK excitations of $pp(p\bar p)$ system
and experimental data.}
\vspace{5mm}
\hbox to \hsize {\hss \small
\begin{tabular}{|c|c|lc|l||c|c|lc|l|}   \hline
n & $M_n^{pp}$\,MeV & $M_{exp}^{pp}$\,MeV & Rfs. & $M_{exp}^{p\bar
p}$\,MeV & n &
$M_n^{pp}$\,MeV & $M_{exp}^{pp}$\,MeV & Rfs. & $M_{exp}^{p\bar
p}$\,MeV \\ \hline
1 & 1878.38 & 1877.5 $\pm$ 0.5 & [9] & 1873 $\pm$ 2.5 & 15 & 2251.68
& 2240 $\pm$ 5
& [13] & 2250 $\pm$ 15   \\ \hline
2 & 1883.87 & 1886 $\pm$ 1 & [6] & 1870 $\pm$ 10 & 16 & 2298.57 &
2282 $\pm$ 4 &
[6,14] & 2300 $\pm$ 20  \\ \hline 
3 & 1892.98 & 1898 $\pm$ 1 & [6] & 1897 $\pm$ 1 & 17 & 2347.45 &
2350
& [15] & 2340 $\pm$ 40 \\ \hline 
4 & 1905.66 & 1904 $\pm$ 2 & [10] & 1910 $\pm$ 30  & 18 & 2398.21 &
&  & 2380 $\pm$ 10
\\ \hline 
5 & 1921.84 & 1916 $\pm$ 2 & [6] & $\sim
$ 1920 &  19 & 2450.73 &   &
& 2450
$\pm$ 10   \\ \cline{6-10}       
  &         & 1926 $\pm$ 2 & [10] &   & 20 & 2504.90 &   &  & $\sim$
  2500    \\ \hline 
  &         & 1937 $\pm$ 2 & [6] & 1939 $\pm$ 2 & 21 & 2560.61 &   &
  &  \\ \cline{6-10} 
6 & 1941.44 & 1942 $\pm$ 2 & [10] & 1940 $\pm$ 1 & 22 & 2617.76 &  &
& $\sim$ 2620
  \\ \cline{6-10}
  &         & $\sim$1945   & [7] & 1942 $\pm$ 5 & 23 & 2676.27 &
  & &          \\ \hline
7 & 1964.35 & 1965 $\pm$ 2 & [6] & 1968 & 24 & 2736.04 & 2735 & [8]
& 2710 $\pm$ 20  \\ \cline{6-10}
  &         & 1969 $\pm$ 2  & [11] & 1960 $\pm$ 15 & 25 & 2796.99 &
  & &
  \\ \hline  
8 & 1990.46 & 1980 $\pm$ 2 & [6] & 1990 $\frac{+15}{-30}$ & 26 &
2859.05 &   & & 2850 $\pm$
5         \\ \cline{6-10}
  &         & 1999 $\pm$ 2 & [6] &  &  27 & 2922.15 &  & &
\\ \hline
9 & 2019.63 & 2017 $\pm$ 3 & [6] & 2020 $\pm$ 3 & 28 & 2986.22 &   &
&   \\
\hline 
10 & 2051.75 & 2046 $\pm$ 3 & [6] & 2040 $\pm$ 40  & 29 & 3051.20 &
& &
   \\ \cline{6-10}
   &         & $\sim$2050   & [12] & 2060 $\pm$ 20 &  30 & 3117.04
   &  &
&  \\ \hline
11 & 2086.68 & 2087 $\pm$ 3 & [6] & 2080 $\pm$ 10 &  31 & 3183.67 &
& &  \\ \cline{6-10}
   &         &              &     & 2090 $\pm$ 20 & 32 & 3251.06 &
   & &            \\ \hline 
   &         & 2120 $\pm$ 3.2 & [12] & 2105 $\pm$ 15 & 33 & 3319.15 &
   & & \\ \cline{6-10}
12 & 2124.27 & 2121 $\pm$ 3 & [13] & 2110 $\pm$ 10 & 34 & 3387.90 & &
& 3370 $\pm$
10 \\ \cline{6-10} 
  &         & 2129 $\pm$ 5 & [6] & 2140 $\pm$ 30 & 35 & 3457.28
& & & \\ \hline
13 & 2164.39 & 2150 $\pm$ 12.6 & [12] & 2165 $\pm$ 45 & 36 & 3527.25
&  & &  \\ \cline{6-10} 
   &         & 2172 $\pm$ 5 & [6] & 2180 $\pm$ 10 & 37 & 3597.77 &  &
   & 3600
$\pm$ 20         \\ \hline 
14 & 2206.91 & 2192 $\pm$ 3  & [13] & 2207 $\pm$ 13 & 38 & 3668.81 &
& &  \\ \hline 
\end{tabular}
\hss}
\end{center}
\end{table}
We have used Review of Particle Physics \cite{4} and recent review
article of Crystal Barrel Collaboration \cite{5} where the
experimental data on mass spectrum of the resonance states of
proton-antiproton system above elastic threshold have been extracted
from. As it is seen from Table 1, the nucleon-nucleon dynamics at low
energies provides a quite remarkable confirmation of Kaluza-Klein
picture. Moreover, Kaluza-Klein scenario predict a special sort of
(super)symmetry between fermionic (dibaryon) and bosonic states,
which is very nontrivial fact, and Table 1 contains an experimental
confirmation of this fact as well. 

Actually, we also see that there are an empty cells, especially
$M_{21}$, $M_{23}$, $M_{25}$, $M_{27}$ -- $M_{33}$, $M_{35}$,
$M_{36}$, in the Table. In this respect we would like to request
nuclear and particle physicists-experimenters to search missing
two-nucleon states to fill the empty cells. We very hope that it
would be possible to make it in the near future. For a convenience we
also present here all possible of the calculated  two-nucleon KK
excitations with account of proton-neutron mass difference. There are
all collected in Table 2, which may serve as a guide for the
experimenters.  

It should be stressed that we have not seen proton-antiproton
resonances in the structure of proton-antiproton total cross section
at low energies. This fact could be explained by crossing properties
of the amplitudes. The crossing structure of the amplitudes is such
as to result to suppression of nucleon-nucleon KK excitations in
proton-antiproton channel and vice versa there is their enhancement
in proton-proton channel.

\section{Conclusion}

In this note we have presented additional arguments in favour of
Kaluza and Klein picture of the world. We have shown that geniusly
simple formula (\ref{KK}) provided by Kaluza-Klein approach gives an
excellent description for the mass spectrum of two-nucleon system.
It has also been established that the experimental data obtained at
low energies where the nucleon-nucleon dynamics has been studied
reveal a special sort of (super)symmetry between fermionic (dibaryon)
and bosonic states predicted by Kaluza-Klein scenario. Of course it
would be very desirable to state new experiments such as e.g. to fill
an empty cells in Table 1 (like in Mendeleev Table!), and we hope
this is a quite promising subject of investigations in particle and
nuclear physics.

\begin{table}[tbp]
\begin{center}
\caption{Kaluza-Klein tower of KK excitations of two-nucleon system.}
\vspace{5mm}
\begin{tabular}{|c|c|c|c|}   \hline
n & $M_n^{pp}$\,MeV & $M_n^{pn}$\,MeV & $M_n^{nn}$\,MeV  \\ \hline
1  & 1878.38 & 1879.67 & 1880.96 \\
2  & 1883.87 & 1885.15 & 1886.44 \\
3  & 1892.98 & 1894.26 & 1895.54 \\
4  & 1905.66 & 1906.93 & 1908.21 \\
5  & 1921.84 & 1923.11 & 1924.37 \\
6  & 1941.44 & 1942.69 & 1943.94 \\
7  & 1964.35 & 1965.59 & 1966.82 \\
8  & 1990.46 & 1991.68 & 1992.89 \\
9  & 2019.63 & 2020.84 & 2022.04 \\
10 & 2051.75 & 2052.94 & 2054.12 \\
11 & 2086.68 & 2087.84 & 2089.01 \\ 
12 & 2124.27 & 2125.42 & 2126.56 \\
13 & 2164.39 & 2165.52 & 2166.64 \\
14 & 2206.91 & 2208.01 & 2209.11 \\
15 & 2251.68 & 2252.75 & 2253.83 \\
16 & 2298.57 & 2299.62 & 2300.68 \\
17 & 2347.45 & 2348.49 & 2349.52 \\
18 & 2398.21 & 2399.23 & 2400.24 \\
19 & 2450.73 & 2451.72 & 2452.72 \\
20 & 2504.90 & 2505.87 & 2506.84 \\
21 & 2560.61 & 2561.56 & 2562.51 \\
22 & 2617.76 & 2618.69 & 2619.62 \\
23 & 2676.27 & 2677.17 & 2678.08 \\
24 & 2736.04 & 2736.92 & 2737.81 \\
25 & 2796.99 & 2797.86 & 2798.73 \\
26 & 2859.05 & 2859.90 & 2860.75 \\
27 & 2922.15 & 2922.98 & 2923.81 \\
28 & 2986.22 & 2987.03 & 2987.85 \\
29 & 3051.20 & 3052.00 & 3052.79 \\
30 & 3117.04 & 3117.81 & 3118.59 \\
31 & 3183.67 & 3184.43 & 3185.20 \\
32 & 3251.06 & 3251.80 & 3252.55 \\
33 & 3319.15 & 3319.88 & 3320.61 \\
34 & 3387.90 & 3388.62 & 3389.34 \\
35 & 3457.28 & 3457.99 & 3458.69 \\
36 & 3527.25 & 3527.94 & 3528.63 \\
37 & 3597.77 & 3598.44 & 3599.12 \\
38 & 3668.81 & 3669.47 & 3670.13 \\ \hline
\end{tabular}
\end{center}
\end{table}

\end{document}